\documentclass{aa}  

\usepackage{graphicx}
\usepackage{titlesec}
\newcounter{subsubsubsection}[subsubsection]

\usepackage{txfonts}
\usepackage{hyperref}
\hypersetup{colorlinks=true,linkcolor=blue,citecolor=blue,filecolor=blue,urlcolor=blue}
\begin{document}

   \title{Revisiting the fundamental parameters for the black hole X-ray transient Swift J1753.5-0127}

   \author{I. V. Yanes-Rizo
          \inst{1, 2}\fnmsep\thanks{\email{idairayr@iac.es}}
          \and
          M. A. P. Torres\inst{1,2}
          \and
          J. Casares\inst{1,2}
          \and
          P. G. Jonker\inst{3,4}
          \and 
          J. Sánchez-Sierras\inst{1,2}
          \and
          T. Muñoz-Darias\inst{1,2}
          \and
          M. Armas Padilla\inst{1,2}
          }

   \institute{Instituto de Astrofísica de Canarias, E-38205 La Laguna, S/C de Tenerife, Spain\\
         \and 
             Departamento de Astrofísica, Universidad de La Laguna, E-38206 La Laguna, S/C de Tenerife, Spain\\
         \and
             Department of Astrophysics\,/\,IMAPP, Radboud University, Heyendaalseweg 135, NL-6525 AJ Nijmegen\\
         \and
             SRON, Netherlands Institute for Space Research, Niels Bohrweg 4, 2333 CA, Leiden, The Netherlands\\
             }

   \date{Received ZZZ; accepted ZZZ}

  \abstract
   {We present time-resolved Gran Telescopio Canarias optical spectroscopy and William Herschel Telescope $i$-band photometry of the X-ray transient SWIFT J1753.5-0127 in quiescence. The $i$-band light curve is dominated by flickering with an amplitude of $\sim 0.5$\,mag and shows no evidence of the ellipsoidal modulation of the companion star. The telluric-corrected average spectrum, on the other hand, reveals the presence of weak (strongly veiled) TiO bands at $7055$\,\AA\,and $7589$\,\AA\,. We used them for a spectral classification, finding an M4-5 V companion star. However, as velocity shifts are not clearly detected in the individual spectra, we turned the analysis to the double-peaked H$\alpha$ emission line from the accretion disc. By exploiting the empirical correlations established for quiescent X-ray transients between the line morphology and fundamental binary parameters, we estimated the radial velocity semi-amplitude of the companion $K_2 = 820 \pm 36$\,km s$^{-1}$, a mass ratio $q = 0.023 \pm 0.006$ and an inclination $i = 79 \pm 5$\,deg. Moreover, an orbital period of $3.26 \pm 0.02$\,h was measured from the modulation of the centroid velocities and the double-peak trough depth of the H$\alpha$ profile. These quantities yielded a mass function $f(M_1) = 7.8 \pm 1.0$\,M$_\odot$ and black hole and companion star masses of $M_1 = 8.8 \pm 1.3$\,M$_\odot$ and $M_2 = 0.20 \pm 0.06$\,M$_\odot$, respectively. The companion star mass is in line with the spectral classification obtained from the relative depth of the TiO bands. Based on the mean quiescent magnitude ($i = 21.4 \pm 0.1$), orbital period, and interstellar extinction, we estimate the distance to the source to be $3.9 \pm 0.7$\,kpc and a Galactic plane elevation of $0.8 \pm 0.2$\,kpc, supporting the case for a large natal kick. }


   \maketitle

\section{Introduction}
Stellar mass black holes (BHs) in the Galaxy are mostly identified in X-ray transients (XRTs), a sub-class of low mass X-ray binaries where a compact object accretes material from a $\leq 1$\,M$_\odot$ stellar companion. They are discovered through dramatic X-ray outburst that recur with typical time scales of decades to centuries \citep{mcclintock2006}. To date, 20 BHs have been dynamically confirmed among the 72 candidate BH XRTs discovered \citep{corral2016}. These dynamical studies rely heavily on the detection of the companion star \citep[see][for a review of dynamical mass measurements in BH XRTs]{casares2014}. In many cases, the stellar companion may not be visible due to the faintness of the optical counterpart and the contaminating radiation from an accretion disc, a jet, or a line-of-sight interloper \citep[e.g. see][]{torres2015, yanes2024}. With the aim of improving the statistics of BH masses, indirect techniques have been developed based on a sample of well-characterised dynamical BH XRTs. Empirical relations have been established to estimate the fundamental parameters of the binaries by exploiting the double-peaked H$\alpha$ emission line profile that arises from the accretion disc around the compact star (\citealt{casares2015, Casares2016, casares2018, casares2022}; see also \citealt{cuneo2023} for an extension to the near infrared). In this work, we employ these tools in order to fully characterize the XRT SWIFT J1753.5-0127 (hereafter J1753).

The XRT J1753 was discovered by the Swift Burst Alert Telescope \citep{barthelmy2005} in May 2005 \citep{palmer2005}. The source was soon detected in other energy bands such as ultraviolet \citep{still2005} and radio \citep{fender2005}. The optical and near-infrared counterparts had peak magnitudes at $R \sim 15.8$, $K = 14.3$ \citep{halpern2005, torres2005}. In the initial 100 days, the X-ray source displayed a fast rise and an exponential decay light curve. Thereafter, it remained in a standstill active state for over 12 years \citep[see e.g.][]{yang2022}. For most of this long-lived outburst, J1753 stayed in the low-hard state, showing a number of spectral softening episodes \citep{soleri2013, yoshikawa2015, shaw2016}. He\,{\sc ii} 4686\,\AA\ and H$\alpha$ emissions were observed during the first days of the eruption \citep{torres2005, torres2005b}. Both lines, typically seen in active BH XRTs, were not apparent a month later \citep{cadolle2007}, and the spectrum remained relatively featureless in subsequent years \citep{jonker2008, durant2009}. By 2009, the only well-constrained binary parameter was the orbital period. A superhump period of $3.2443 \pm 0.0010$\,h was reported by \citet{zurita2008} based on $R$-band light curves. The superhump modulation is attributed to a slowly precessing elliptical accretion disc, a scenario in which the orbital period is shorter by a few percent than the value established for the superhump. A subsequent photometric study performed by \citet{paez2012} confirmed this period and refined the value to $3.2448 \pm 0.0002$\,h. This result was challenged by \citet{neustroev2014}, who proposed an orbital period of $2.85$\,h and the presence of a <5 M$_\odot$ BH in J1753. However, \citet{shaw2016} were not able to replicate the results of \citet{neustroev2014} and derived a lower limit to the compact object mass of $7.4 \pm 1.2$\,M$_\odot$ based on outburst H$\alpha$ spectroscopy. After 12 years of outburst activity, J1753 started fading in September 2016 \citep{russel2016}, although it still exhibited rebrightening episodes in February 2017 \citep{alqasim2017} and April 2017 \citep{bernardini2017} before returning to full quiescence in July 2017 \citep{zhang2017a}. Optical spectroscopy obtained when the source was in early quiescence revealed only a broad H$\alpha$ emission-full width half maximum (FWHM)$\, = 3800$\,km s$^{-1}$, EW$\, = 90$\,\AA\,\citep{neustroev2017}. SWIFT J1753.5-0127 was recently found to be active again \citep{alabarta2023}, although the new outburst only lasted for five months and the binary has been back to the quiescent state since September 2024 \citep{alabartarus2024}.

Studies of the optical counterpart have been limited due to a lack of emission features during extended periods of the outburst phase and its faint magnitude during quiescence. Thus, some of the most fundamental properties of J1753 remain uncertain. In this context, the orbital inclination is loosely constrained between $40 - 80$\,deg \citep[e.g.][]{neustroev2014, veledina2015}, with the distance estimated to range from 2.5 to 20\,kpc, accounting for uncertainties (see Sect.~\ref{sec:distanceandveiling} for details). In this work, we present time-resolved observations of J1753 conducted in July 2018 during quiescence. The paper is structured as follows: In Sect.~\ref{sec:2}, we detail the observations and the data reduction methods. The results are presented in Sect.~\ref{sec:3}, where we first perform a spectral classification of the companion star using the telluric-corrected average spectrum of J1753. Next, we derive the radial velocity semi-amplitude of the companion star, the mass ratio, the orbital inclination, and the binary period by exploiting the double-peaked H$\alpha$ line properties. Finally, in Sect.~\ref{sec:4}, we establish the masses for the BH and the companion star and further discuss its spectral type. We also estimate the distance to the source, supporting the scenario of a large natal kick.


\section{Observations and data reduction}
\label{sec:2}
\subsection{Spectroscopy}
\label{sec:spectroscopy}
Time-resolved spectroscopy of J1753 was obtained using the OSIRIS spectrograph \citep{cepa2000} mounted on the 10.4-m Gran Telescopio Canarias (GTC) at the Observatorio del Roque de los Muchachos (ORM) on the island of La Palma. A total of 14, 14, and 12 spectra, with an exposure time of 900 s each, were taken on the nights of July 12, 13, and 14, 2018, respectively. The time series spectroscopy cover between 4.3\,h (first night), 4.0\,h (second night), and 3.3\,h (last night). The target was observed at airmass <1.4\,arcsec, and the image quality varied between 1.0 and 1.7 arcsec throughout the observations, with a mean value of 1.4 arcsec. The R1000R grism and a $1.0$ arcsec-long slit were employed with the Marconi CCD44-82 set to a 2x2 binning mode. This configuration enabled us to observe the spectral range $5100-10000$\,\AA\,with a 2.62\,\AA\ pixel$^{-1}$ dispersion. The object was frequently reacquired to have the slit at the parallactic angle in order to minimize flux losses across the spectrum caused by atmospheric refraction. The instrumental setup delivered spectra with a resolution of 7.3\,\AA\,FWHM. 

For reduction, extraction, and wavelength calibration of the data, {\sc pyraf}\footnote{\url{https://github.com/iraf-community/pyraf}} standard routines were used. The pixel-to-wavelength scale for the spectra of J1753 was determined using a two-piece cubic spline fit to ten lines from combined He+Ne arc lamp exposures obtained at the end of each night. The rms scatter of the fit was $< 0.03$\,\AA. Wavelength zero point deviations were calculated by fitting the [O\,{\sc i}] 6300.3\,\AA\ sky line. A signal-to-noise ratio per pixel S/N $\sim$ 5 was measured in the $6000-6200$\,\AA\ wavelength region of the individually extracted spectra.

The data were imported to the {\sc molly}\footnote{\textsc{molly} was written by T.~R. Marsh and is available from \url{https://cygnus.astro.warwick.ac.uk/phsaap/software/molly/html/INDEX.html}.} package, where we applied the wavelength zero point and heliocentric corrections. The spectra were also resampled to a uniform velocity scale of 116\,km s$^{-1}$ and corrected for the instrumental response using the flux standard Ross 640. Since we acquired our target in the $i$-band, the resulting spectral continuum is reliable in this spectral range, except beyond 8200\,\AA\,because of contamination by residual OH sky emission. The individual spectra were corrected for telluric absorption lines using {\sc molecfit} v.3.0.3 \citep{smette2015} following the  method described in \citet{sanchez2023}. Essentially, we selected spectral regions covering the H$_2$O and O$_2$ absorption features, excluding wavelength intervals known to potentially contain characteristic late M dwarf TiO bands. The regions selected are $6766-6966$, $7408-7945$, and $7955-8387$\,\AA. We started the telluric correction by fitting each molecule separately, taking their resulting column densities as initial values, and subsequently simultaneously fit these bands again.

In addition, for the analysis presented in in the next section, we also used flux-calibrated and telluric-corrected M dwarf spectra from the X-shooter library Data Release 3 \citep{verro2022}. These are HD 115404B (M1 V), L 422-13 (M3 V), HD 125455B (M4 V), * omi02 Eri C (M4.5 V), and LP 731-58 (M6.5 V). The X-shooter templates were loaded into {\sc molly} and rebinned to match the same velocity scale as the target data.

\subsection{Photometry}
\label{sec:photometry}
Time-resolved photometry of J1753 was performed with the Auxiliary-port CAMera \citep[ACAM;][]{benn2008} installed on the 4.2-m William Herschel Telescope (WHT) at the ORM. A total of 116 300-s images were taken simultaneously with the spectroscopy data on the nights of 12 and 13 July, 2018, covering 5.3\,h per night. Their bias subtraction and flat field correction were conducted using the {\sc pyraf} software. Optimal aperture photometry \citep{Naylor1998} on J1753 and eight comparison stars was carried out using the HiPERCAM\footnote{\url{https://github.com/HiPERCAM/hipercam}} data reduction pipeline. The comparison stars were used for differential photometry and to verify the resulting photometric variability of J1753. The instrumental magnitudes were calibrated using the mean apparent magnitudes from the Pan-STARSS Data Release 2 catalogue \citep{chambers2016}. 

\section{Analysis and results}
\label{sec:3}

\subsection{Light curves}
\begin{figure}
	\includegraphics[width=\columnwidth]{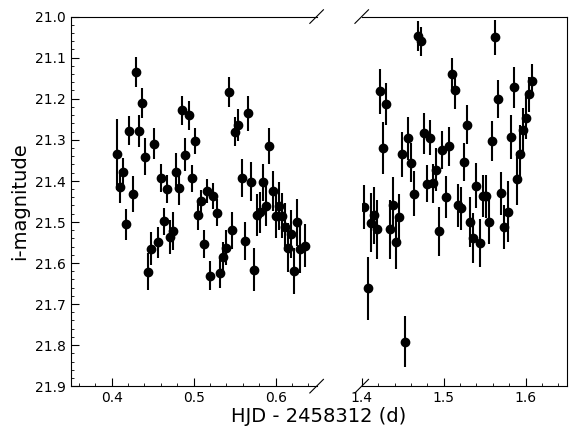}
    \caption{Nightly $i$-band light curves of J1753.}
    \label{fig:lightcurve}
\end{figure}

The $i$-band light curves of J1753 are shown in Fig.~\ref{fig:lightcurve}. We established a mean magnitude of $i = 21.4 \pm 0.1$, where the error comes from the rms variability. This is consistent with the $i = 21.37 \pm 0.28$\,quiescent magnitude reported for the source \citep{alabarta2024}. On the other hand, the rms variability is a factor of two larger than that of field stars of a similar brightness. A Lomb-Scargle periodogram of the data shows no prominent peaks, including at the frequencies associated with the orbital periods proposed by \citet{zurita2008} and \citet{neustroev2014}. This is expected if the variability is dominated by flickering that hampers the potential detection of the 
ellipsoidal modulation associated with the companion (see Sect.~\ref{sec:distanceandveiling} for further details). We estimated a flickering peak-to-peak amplitude of $\sim 0.5$\,mag. For comparison, flickering amplitudes of $\sim$ 0.6 and $\sim$ 1.2 mag were observed in the short period BH XRTs MAXI J1659-152 \citep{corral2018} and SWIFT J1357.2-0933 \citep{shahbaz2013}, respectively.

\subsection{The telluric-corrected average spectrum}
\label{sec:3.1}
\label{sec:averagespectralfeatures}
\begin{figure}
	\includegraphics[width=\columnwidth]{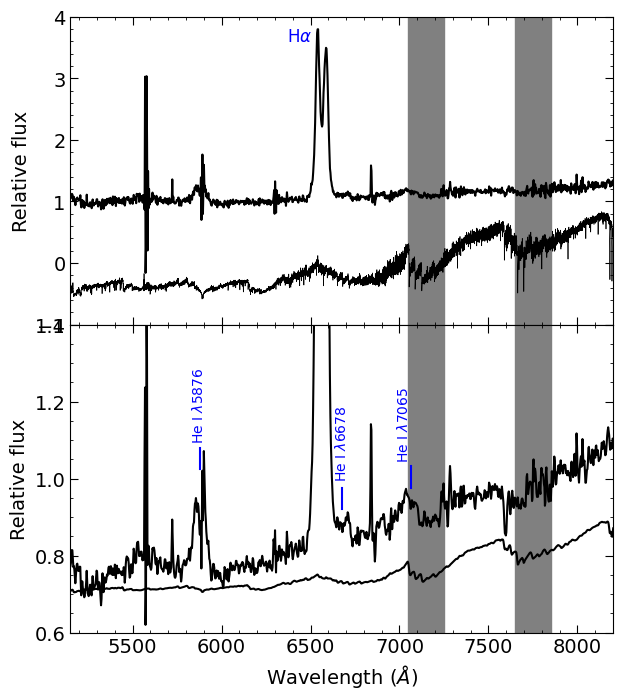}
    \caption{OSIRIS J1753 averaged spectrum. The upper panel displays the telluric-corrected and dereddened spectrum normalised to unity at 7500\AA. The location of the expected M-type companion star TiO molecular bands are marked with grey bands. The lower panel presents a close up view of the same spectrum after smoothing using a Gaussian with FWHM = 2\,pixels. The He\,{\sc i} 5876\,\AA\,, 6678\,\AA\,, and 7065\,\AA\, emission lines are highlighted in blue. An M4 V template star (HD 125455B) broadened to match the target spectral resolution is overplotted. The latter has been scaled by a factor of 0.08 to simulate the companion flux contribution in the $i$-band and shifted in the y-axis for display purposes.}
    \label{fig:promedio_m4v}
\end{figure}

First of all, we dereddened the average telluric-corrected and flux calibrated spectrum of J1753. We adopted an interstellar extinction $E(B-V) = 0.45 \pm 0.09$ from fits to the 2175\,\AA\ absorption feature (\citealt{froning2014}; see also \citealt{rahoui2015} for a coincident estimate using interstellar bands). The resulting spectrum is dominated by a strong ($EW = 166 \pm 16$\,\AA \footnote{The error reflects the standard deviation in the distribution of EW values measured from the individual H$\alpha$ profiles.}) broad and double peaked H$\alpha$ emission as well as weaker He\,{\sc i} 5876\,\AA\ also in emission. An even weaker emission from He\,{\sc i} 6678\,\AA\ (and perhaps He\,{\sc i} 7065\,\AA) was also detected, peeping out over the redwing of H$\alpha$ (see Fig.~\ref{fig:promedio_m4v}). We note that the centroid of the H$\alpha$ line is redshifted by $150$\,km s$^{-1}$.

\begin{figure}
	\includegraphics[width=\columnwidth]{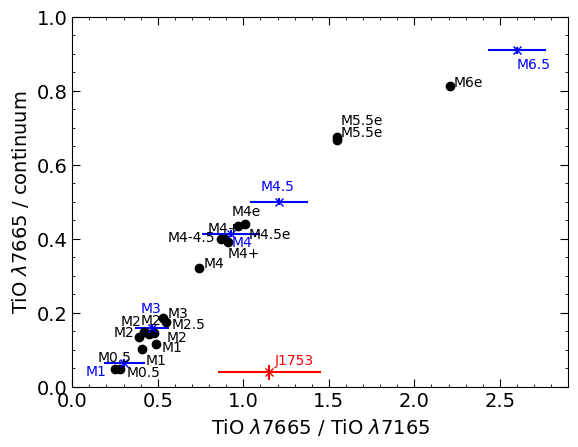}
    \caption{Fractional depth of the TiO band at 7665\,\AA\,as a function of the TiO band ratio $\lambda7165/\lambda7665$ for J1753 and several M dwarf standards from Table 3 in \citet[][black]{wade1988} and the X-shooter spectral library (blue). The TiO band ratio implies a spectral type M4-5 V for the companion star in J1753, where the strength of the 7665\,\AA\,band is reduced due to disc dilution.}
    \label{fig:wadehornefigure}
\end{figure}

A closer inspection of the continuum region revealed weak absorption structures that are consistent with TiO bands at 7055\,\AA\,and 7589\,\AA\,(see Fig.~\ref{fig:promedio_m4v}). We therefore tried to estimate the companion spectral type in J1753 by quantifying the flux deficits across these TiO bands. Following \citet{wade1988} and \citet{marsh1999}, we established the continuum by a linear fit between 7458-7547\,\AA\,, 8127-8167\,\AA\,, and 8242-8278\,\AA, obtaining the flux drop between 7140-7190\,\AA\,for the first TiO band and 7640-7690\,\AA\,for the second. As shown in Fig.~\ref{fig:wadehornefigure}, this result is consistent with an M4-5 V star according to the TiO band ratios for M dwarfs given in Table 3 of \citet{wade1988}. As a test, we also measured and show in Fig.~\ref{fig:wadehornefigure} this ratio for a sample of X-shooter library spectra with reliable spectral classification (Sect.~\ref{sec:spectroscopy}). In addition, we inspected the impact of orbital smearing (assuming the donor's velocity amplitude derived in Sect.~\ref{ref:sect331}) in the calculation of the TiO band ratio and found that it is smaller than the range implied by the uncertainty in our spectral type determination.

Next, we attempted to constrain the companion fractional flux contribution ($f$) to the $i$-band by decomposing the spectrum into its disc and stellar components. For this purpose, we used the M4 V X-shooter spectrum (HD 125455B) previously degraded with a Gaussian with 7.3\,\AA\,FWHM to match the resolution of the OSIRIS data. In addition, the continuum was rectified to unity in both target and template in the $6950-7050$, $7450-7550$, and $8130-8170$\,\AA\, wavelength regions. Various fractions of the normalised template were subtracted from the target spectrum, and the value of the fractional contribution $f$ was optimised in order to find the minimum $\chi^2$ occurring at $f = 0.08 \pm 0.01$. This value must be treated as a lower limit since the J1753 average spectrum was obtained in the rest frame of the observer is therefore velocity smeared. That said, in Sect.~\ref{sec:distanceandveiling} we infer $f < 0.15$ from photometry, which is in good agreement with the above constraint. We attempted to measure radial velocities from the companion star by cross-correlating every individual spectrum with the X-shooter M-type templates in the regions of the TiO bands ($7000-7450$ and $7550-8130$\,\AA) and the vicinity of H$\alpha$ ($6000-6250$ and $6350-6500$\,\AA), but, most of the cross-correlation functions are unfortunately too noisy and do not show significant peaks.

\subsection{Constraints to the binary parameters from the H$\alpha$ emission line}
\label{sec:halfaemissionline}
\begin{figure}
	\includegraphics[width=\columnwidth]{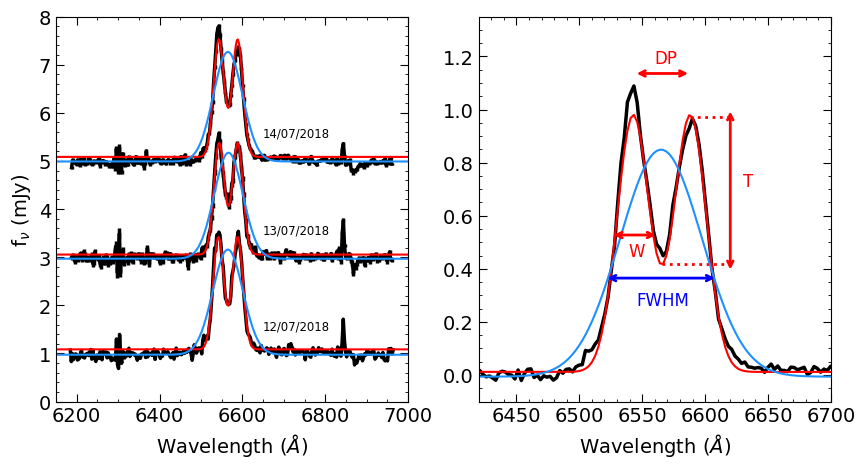}
    \caption{Average H$\alpha$ emission line profiles at the heliocentric rest frame. Spectra have been vertically shifted for display purposes. The blue and red lines represent the best-fitting single and symmetric two-Gaussian models, respectively. The right panel shows the result of combining the total 40 spectra. The relevant parameters for the characterisation of the line profile are marked.}
    \label{fig:orbits}
\end{figure}

Previous analysis of quiescent XRTs has revealed empirical relationships that allow us to estimate the radial velocity semi-amplitude of the companion star ($K_2$), the mass ratio ($q$), and the orbital inclination ($i$) from the morphology of this line \citep[see][for further details]{casares2015, Casares2016, casares2022}. In order to apply these correlations, we first normalised the spectra to the adjacent continuum level. Next, we fit a single and a symmetric two-Gaussian model to the normalised individual data, the average profile for each night, and that obtained from combining all individual data. The models were degraded to the $7.3$\,\AA\ instrumental resolution. From the single Gaussian fit, we measured the FWHM, while from the symmetric two-Gaussian, we measured the FWHM of each Gaussian ($W$) and the Gaussian double-peak separation ($DP$). Table \ref{tab:orbits} shows the profile parameters obtained for each average spectrum. The fits to the night averages allowed us to examine possible orbital cycle-to-cycle line profile variability in our calculations. This variability is illustrated in Fig.~\ref{fig:orbits}, where the three average H$\alpha$ profiles of each night (and their fits) are shown, representing three complete orbits. The double peak becomes increasingly asymmetric on each successive night, with the blue peak exhibiting greater intensity compared to the red peak. These persistent asymmetries have previously been attributed to an eccentric disc with an asymmetric brightness distribution \citep[see][]{zurita2002, torres2004}.

\begin{table*}
	\centering
    \begin{center}
	\caption{H$\alpha$ profile parameters for each night average and the result of combining the full data set.}
	\label{tab:orbits}
	\begin{tabular}{  c  c  c  c  c  c  c c }
		\hline
            & $N^{(a)}$ & $FWHM$ (km s$^{-1}$) & $DP$ (km s$^{-1}$) & $W$ (km s$^{-1}$) & $T$ & $q$ & $i$ (deg) \\ \hline 
		12/07/2018 & 
	    \begin{tabular}{c} $14$ \\
            \end{tabular} & 
	    \begin{tabular}{c} $3589 \pm 30$ \\
            \end{tabular} &
        \begin{tabular}{c} $2119 \pm 9$ \\
            \end{tabular} & 
        \begin{tabular}{c} $1370 \pm 11$ \\
            \end{tabular} & 
        \begin{tabular}{c} $0.62 \pm 0.01$ \\
            \end{tabular} &
        \begin{tabular}{c} $0.027 \pm 0.006$ \\
            \end{tabular} & 
        \begin{tabular}{c} $82 \pm 5$ \\
            \end{tabular} \\ \hline
		13/07/2018 & 
	    \begin{tabular}{c} $14$ \\
            \end{tabular} &
        \begin{tabular}{c} $3414 \pm 26$ \\
            \end{tabular} &
        \begin{tabular}{c} $2049 \pm 9$ \\
            \end{tabular} & 
        \begin{tabular}{c} $1372 \pm 10$ \\
            \end{tabular} &
        \begin{tabular}{c} $0.58 \pm 0.01$ \\
            \end{tabular} &
        \begin{tabular}{c} $0.018 \pm 0.004$ \\
            \end{tabular} & 
        \begin{tabular}{c} $78 \pm 5$ \\
            \end{tabular} \\ \hline
		14/07/2018 & 
	    \begin{tabular}{c} $12$ \\
            \end{tabular} &
        \begin{tabular}{c} $3495 \pm 19$ \\
            \end{tabular} &
        \begin{tabular}{c} $2073 \pm 6$ \\
            \end{tabular} & 
        \begin{tabular}{c} $1372 \pm 7$ \\
            \end{tabular} &
        \begin{tabular}{c} $0.593 \pm 0.008$ \\
            \end{tabular} &
        \begin{tabular}{c} $0.024 \pm 0.003$ \\
            \end{tabular} &
        \begin{tabular}{c} $79 \pm 5$ \\
            \end{tabular} \\ \hline \hline
		All spectra &
	    \begin{tabular}{c} $40$ \\
            \end{tabular} &
        \begin{tabular}{c} $3514 \pm 14$ \\
            \end{tabular} &
        \begin{tabular}{c} $2075 \pm 4$ \\
            \end{tabular} & 
        \begin{tabular}{c} $1374 \pm 5$ \\
            \end{tabular} &
        \begin{tabular}{c} $0.593 \pm 0.005$ \\
            \end{tabular} & 
        \begin{tabular}{c} $0.027 \pm 0.003$ \\
            \end{tabular} & 
        \begin{tabular}{c} $79 \pm 5$ \\
            \end{tabular} \\ \hline
	\end{tabular}    
    \end{center}
    \tablefoot{$^{(a)}N$: Number of spectra that have been averaged.\\
    All quoted uncertainties are at 68\% confidence level.}
\end{table*}

\subsubsection{The radial velocity semi-amplitude of the companion star $K_2$, mass ratio $q$ and orbital inclination $i$}
\label{ref:sect331}

The first relation we exploited, presented in \citet{casares2015}, links $K_2$ with the FWHM of the H$\alpha$ emission line by following
\begin{equation}
    K_2 = 0.233(13) \times \rm{FWHM}~.
	\label{eq:K2}
\end{equation}

In accordance with the procedure applied in \citet{casares2015}, we obtained FWHM$ = 3519 \pm 155$\,km s$^{-1}$ from the Gaussian fits to the individual profiles, where the uncertainty corresponds to the standard deviation. This yielded $K_2 = 820 \pm 36$\,km s$^{-1}$. In our determination, we have assumed that the above linear relation can be extrapolated from the upper FWHM value of $2850$\,km s$^{-1}$ \citep{casares2015} to a FWHM of $3519$\,km s$^{-1}$ as measured for J1753.

On the other hand, \citet{Casares2016} have presented a correlation between $q$ and the ratio of the double-peak separation, $DP$, to the FWHM of the H$\alpha$ line, given by 
\begin{equation}
    \log q = -6.88(0.52) - 23.2(2.0) \log\left(\frac{DP}{FWHM}\right)~.
	\label{eq:massratio}
\end{equation}
Table \ref{tab:orbits} lists the values of $DP$ and FWHM obtained for each nightly averaged profile and the combined average profile. The table also shows the values of $q$ that were computed through a Monte Carlo randomisation with 10 0000 realisations. All the quoted uncertainties correspond to the 68\% confidence level. From the full averaged profile, we obtained a mass ratio of $0.027 \pm 0.003$. We also established $q = 0.023 \pm 0.006$ by combining the normal distributions from each night. Both values are consistent within 68$\%$, and we took the latter value, as it is the most conservative estimate.

The orbital inclination $i$ can be obtained by using the correlation developed by \citet{casares2022} that links this parameter with the depth of the trough ($T$) between the two peaks of the H$\alpha$ emission profile following
\begin{equation}
    i \,(\rm{deg}) = 93.5(6.5)\,T + 23.7(2.5)~,
	\label{eq:inclination}
\end{equation}
where $T$ is given by
\begin{equation}
    T = 1 - 2 ^{1-\left(\frac{DP}{W}\right)^2}~.
	\label{eq:depth}
\end{equation}

Table \ref{tab:orbits} provides the parameters $W$ and $DP$ for each averaged profile and the associated $T$. We note that the values of $T$ have been shifted by 0.01 to correct for instrument resolution degradation. This offset was obtained from equation A1 in Appendix A in \citet{casares2022}. The last column of Table \ref{tab:orbits} gives the inclination obtained through Eq.~\ref{eq:inclination} after performing a Monte Carlo simulation with 10 000 steps. We established $i = 79 \pm 5$\,deg (68\% confidence level) by adding the normal distributions for each night. A fully consistent result was obtained from the full averaged profile. We caution that our $i$ determination is based on the assumption that the $T-i$ correlation can be extrapolated to higher inclination values than those where it was obtained (i.e. $ 35 - 75$\,deg). 

\subsubsection{The orbital period}
\begin{figure}
	\includegraphics[width=\columnwidth]{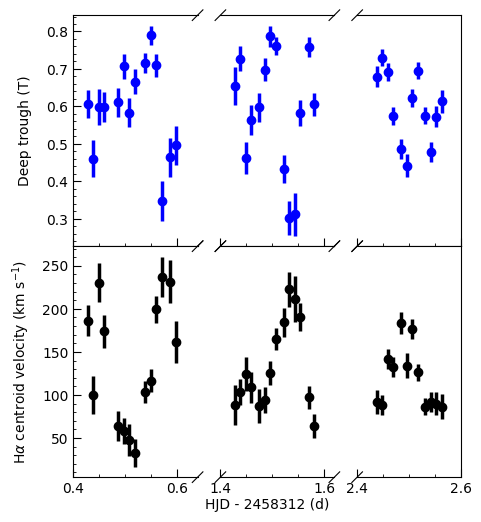}
    \caption{Time evolution of $T$ (top panel) and the H$\alpha$ emission line centroid velocities (bottom) from data obtained on July 12, 13, and 14, 2018.}
    \label{fig:Tandcentroids}
\end{figure}

The time evolution of the individual H$\alpha$ emission profiles is a key tool for tracing the motion of the accretion disc. Specifically, it has been found that in BH XRTs, the depth of the trough ($T$) is modulated with the orbital phase \citep{casares2022}. The $T$ curve displays a double-humped morphology with maxima at phases 0.2 and 0.7 and unequal minima at 0.45 and 0.95, where the absolute phase zero is defined as the inferior conjunction of the secondary star. The phasing is attributed to the periodic motion of S-waves (tied to the hot spot and/or the companion star) across the line profile \citep{casares2022}. Additionally, the minimum at 0.95 appears deeper than the minimum at phase 0.45, probably due to obscuration of the hot spot by the outer disc \citep[see][for further details]{casares2022}. Taking this into account, the modulation in $T$ can be employed to determine the orbital period in systems where the companion star is veiled by the accretion disc, as is the case here. Figure\,~\ref{fig:Tandcentroids} (top panel) displays the time evolution of $T$, where a double-humped modulation can be seen, and it is especially clear on the second night. The centroid velocities of the H$\alpha$ line profiles can also be used to trace the orbital motion of the companion star \citep[e.g.][]{mata2015}. By fitting a symmetric two-Gaussian model to each individual profile, we obtained the radial velocity curve of the H$\alpha$ centroid. This is shown in the bottom panel of Fig.~\ref{fig:Tandcentroids}, where a clear modulation with a $\sim 150$\,km s$^{-1}$ amplitude can also be seen. 

\begin{figure}
	\includegraphics[width=\columnwidth]{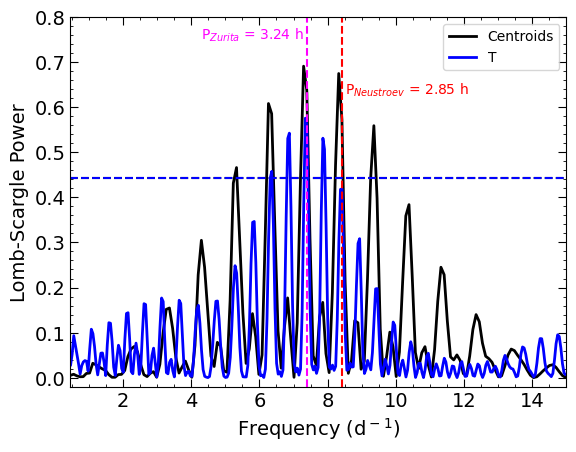}
    \caption{Lomb-Scargle periodograms obtained from the variability of the double peak trough $T$ (blue) and the H$\alpha$ emission line centroid velocities (black). For comparison purposes, and due to the double-humped nature of the $T$ modulation, we halved the frequencies corresponding to the periodogram associated with $T$. The horizontal black line shows the 99\% white noise significance level. The pink and red vertical dashed lines mark the potential orbital periods found by \citet{zurita2008} and \citet{neustroev2014}, respectively.}
    \label{fig:periodogram}
\end{figure}

In order to search for the orbital period, we computed Lomb-Scargle periodograms \citep{lomb1976, scargle1982} for the $T$ and centroid velocity evolution (see Fig.~\ref{fig:periodogram}). Since the $T$ modulation shows a double hump in an orbital cycle, we divided by two the frequencies corresponding to its periodogram. In both cases, the frequency of the highest peak is found at 7.34 cycle d$^{-1}$ (0.136-d period). Fitting a single Gaussian to each peak yielded a period of $3.26 \pm 0.02$\,h ($0.136 \pm 0.001$\,d), where the uncertainty represents the standard deviation of the fit. Since this is the only concurrent peak above the 99\% white noise level, we adopted this spectroscopic determination as the orbital period in J1753. Although additional peaks appear over this threshold, they do not coincide in both periodograms. For example,  the period claimed by \citet{neustroev2014} is only the fifth peak in our $T$ periodogram, and it does not rise above the 99\% level. We interpret it as an alias of the true orbital period.

\begin{figure}
	\includegraphics[width=\columnwidth]{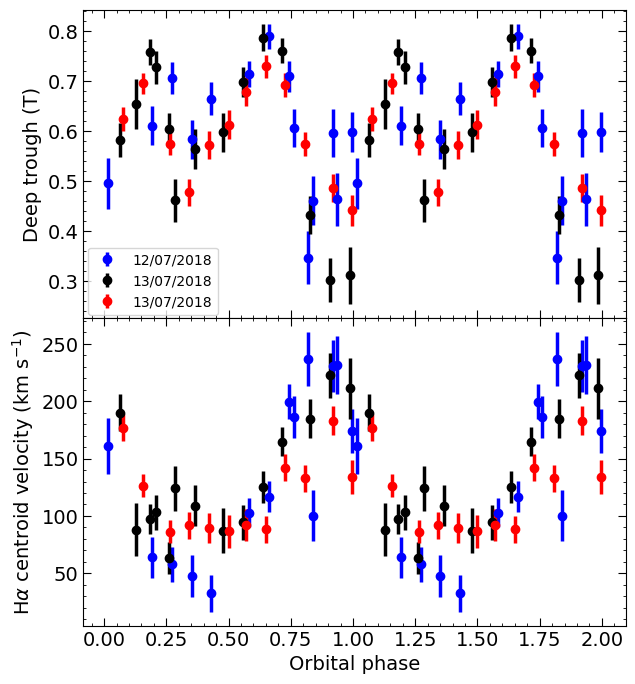}
    \caption{Orbital evolution of $T$ (top panel) and the H$\alpha$ centroid velocities (bottom panel). The curves have been folded using our spectroscopic orbital period $3.26 \pm 0.02$\,h and a tentative $T_0(HJD) = 2458313.41 \pm 0.05$\,d, obtained by imposing that the deepest minima in the depth of the trough $T$ are centred at phase 0.95. For the sake of clarity, one orbital cycle is repeated. The $T$ curve shows a double-humped morphology with the orbital period.}
    \label{fig:phasefold}
\end{figure}

Figure\,~\ref{fig:phasefold} shows the orbital evolution of $T$ and the H$\alpha$ emission line centroid velocities folded on our spectroscopic orbital period ($3.26 \pm 0.02$\,h). The time of phase zero, $T_0$, was tentatively chosen so that the deepest minimum in $T$ occurs at the absolute orbital phase 0.95, as mentioned before. Fitting a Gaussian to the three deepest minima (see Fig.~\ref{fig:phasefold}) yielded $T_{0.95}(HJD) = 2458313.54 \pm 0.05$\,d. This translates into $T_0(HJD) = 2458313.41 \pm 0.05$\,d. With this $T_0$, the phase-folded curve of the H$\alpha$ centroid velocity presents a minimum at phase $\sim 0.4$ and a maximum at phase $\sim 0.9$. We note that this is shifted by +0.15 cycles with respect to the expected radial velocity curve of the compact star, although large phase shifts in emission line velocities caused by contamination from the hot stop are not uncommon \citep[e.g.][]{orosz1994}.

\subsubsection{Skew mapping}
Having determined the orbital period, we could then attempt to constrain the radial velocity semi-amplitude $K_2$ of the companion star using skew-mapping techniques \citep{vandeputte2003}. Essentially, we shifted the 40 individual spectra of J1753 to the donor’s rest frame, assuming a pair of $K_2$ and $T_0$ values. The resulting summed spectrum was cross-correlated with the M4 V template in the same wavelength interval used in Sect.~\ref{sec:averagespectralfeatures} for the spectral classification. We repeated the process for a range of possible $K_2$ and $T_0$ values and the peak of the cross-correlation function was plotted as a function of $K_2$ and $T_0$ in a 2D map. The highest peaks should appear around the correct $K_2$ and $T_0$ values, whereas an incorrect choice of $K_2$ and $T_0$ would result in noisy cross-correlations and thus lower peaks. 

\begin{figure*}
	\centering \includegraphics[height=12cm]{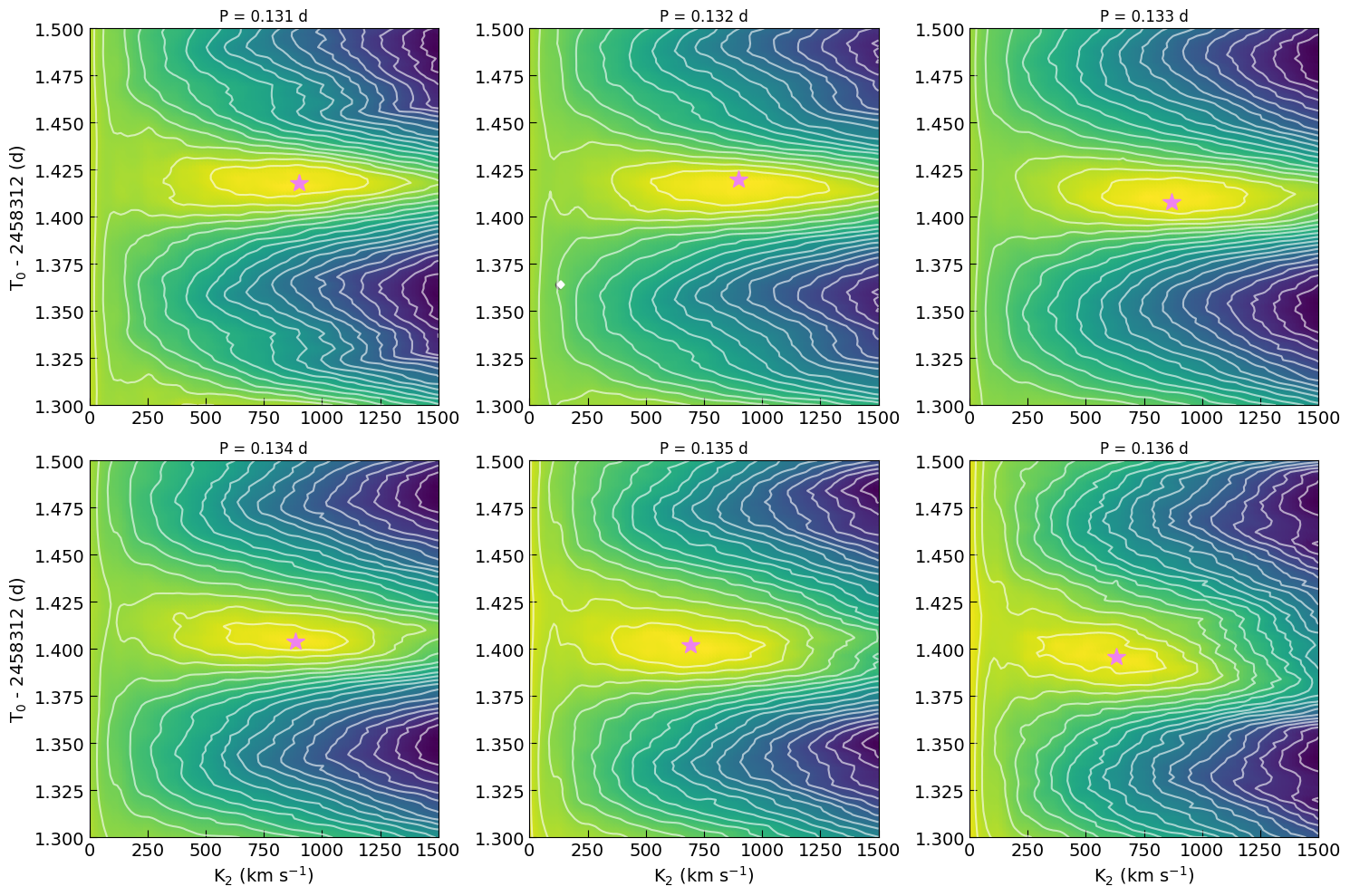}
    \caption{Mosaic of cross-corelation peak values obtained for a range of $K_2$ and $T_0$, and several periods between 0.131 - 0.136 d. The lightest (yellow colour) regions correspond to the highest cross-correlation values, normalised to the peak (marked with a pink star). Each contour is separated by 0.05 in the normalised scale. The highest cross-correlation peaks are constrained between $T_0 (HJD) = 2458313.40 - 2458313.42$\,d and $K_2 = 650-950$\,km s$^{-1}$. For periods outside the plotted range, the contours become noisier and more distorted.}
    \label{fig:skew_map}
\end{figure*}

We produced a grid of cross-correlation functions by searching $K_2$ in the range $0-1500$\,km s$^{-1}$ in 10\,km s$^{-1}$ steps and with $T_0(HJD)$ centred on 2458313.41 d, covering a full orbital cycle with a step size of 0.001 d. We also varied the orbital period between 0.130 and 0.140 d to explore the sensitivity of the $K_2-T_0$ maps to the orbital period. We find that all the maps computed for periods in the range 0.131 - 0.136 d produce clear peaks in the cross-correlation surfaces around the $T_0$ and $K_2$ values expected from our previous analysis, which is reassuring (see Fig.~\ref{fig:skew_map}). In particular, the contour plots show elongated spots that shift between $T_0 (HJD) = 2458313.40 - 2458313.42$ and $K_2 = 650 - 950$\,km s$^{-1}$, depending on the choice of orbital period. While the $T_0$ values of the cross-correlation peaks are in excellent agreement with our previous $T_0$ estimate, the elongation of the contours in the horizontal axis allow for a wider range of $K_2$ values, and therefore this parameter is not well constrained. Consequently, in the remainder of this paper we adopt $K_2 = 820 \pm 36$\,km s$^{-1}$, as derived from the $FWHM-K2$ correlation.

\section{Discussion}
\label{sec:4}
Previous studies have attempted to calculate the system parameters in J1753. However, this has proved challenging due to its prolonged active state (more than 11 years) and the faintness of the optical counterpart in quiescence. Our GTC campaign, obtained during the quiescent period, shows a prominent double-peaked H$\alpha$ emission line that we exploit to constrain the fundamental binary parameters. We derived $K_2 = 820 \pm 36$\,km s$^{-1}$, which is consistent with the lower limit found by \citet{shaw2016} during outburst ($K_2 = 692 \pm 20$\,km s$^{-1}$). We also obtained $q = 0.023 \pm 0.006$, which compares well with $q \sim 0.025$ as estimated by \citet{zurita2008} from their superhump and disc precession periods. A period analysis of the $T$ and centroid velocity curves revealed an orbital period of $3.26 \pm 0.02$\,h. This is consistent with the superhump period of \citet{zurita2008} and \citet{paez2012}. On the other hand, the orbital inclination was loosely constrained in the literature. In this work, we derived $i = 79 \pm 5$\,deg based on the $T-i$ correlation. Armed with these quantities, we could then infer the masses of the stellar components in J1753.

\subsection{Black hole and companion star masses}
In order to constrain the BH and companion star masses, we ran a Monte Carlo simulation with 10 000 random sets. The orbital period and the $K_2$ and $q$ values obtained in this paper yield a mass function of $f(M_1) = 7.8 \pm 1.0$\,M$_\odot$. The non-detection of X-ray eclipses during outburst in the published literature of J1753 allowed us to establish a prior on the upper limit to the inclination set by the geometrical constraint: $\cos{i} \geq 0.49q^{2/3}\left[0.6q^{2/3}+ \ln{\left(1+q^{1/3}\right)}\right]^{-1}$. The masses of the BH and its companion star could then be obtained from:
\begin{equation}
    M_1 = \frac{f(M_1)(1+q)^2}{\sin^3{i}}; \; \; \; \; \; M_2 = qM_1~.
    \label{eq:massfun}
\end{equation}

Adding our inclination estimate of $i = 79 \pm 5$\,deg and the geometrical prior for no X-ray eclipses resulted in $M_1 = 8.8 \pm 1.3$\,M$_\odot$ and $M_2 = 0.20 \pm 0.06$\,M$_\odot$, at the 68\% confidence level. The BH mass is in line with the lower limit $7.4 \pm 1.2$\,M$_\odot$ provided by \citet{shaw2016}. The BH mass can also be estimated from the relation between the FWHM of the symmetric two-Gaussian model fit (i.e. $W$) and the orbital period ($P_{\rm{orb}}$) through the equation:
\begin{equation}
    M_1^* = 3.45\times10^{-8} P_{\rm{orb}} \left(\frac{0.63W+145}{0.84}\right)^3   \rm{M_\odot}~,
	\label{eq:mass}
\end{equation}
with $P_{\rm{orb}}$ expressed in days and the associated uncertainty set to 20\% of the resulting value \citep{casares2022}. The combination of the distributions obtained for $M_1^*$ from the nightly measurements of $W$ yielded a mean $M_1^* = 8.1 \pm 1.6$\,M$_\odot$, which is in good agreement with the $M_1$ value provided by Eq.~\ref{eq:massfun}.

The companion star mass that we derive is consistent with a M3-5 V star \citep[$0.37 - 0.162$\,M$_\odot$\footnote{\url{https://www.pas.rochester.edu/~emamajek/EEM_dwarf_UBVIJHK_colors_Teff.txt}};][]{pecaut2013}. This is in excellent agreement with the M4-5 V spectroscopic classification estimated in Sect.~\ref{sec:averagespectralfeatures}. As an independent test, we can constrain its spectral type of a Roche lobe filling star using the relation $\bar{\rho} \approx 110 \times P^{-2}_{\rm{orb}}$\,gr cm$^{-3}$, where $\bar{\rho}$ is the mean stellar density and $P_{\rm{orb}}$ is the orbital period in hours \citep{frank2002}. Our orbital period implies a stellar density of $\simeq10.4$\,gr cm$^{-3}$, which compares well with the density $11.1$\,gr cm$^{-3}$ of an M3 V star.\footnote{We computed this density value using the stellar mass and radius reported online by \citet{pecaut2013}.}.

\subsection{The distance, stellar contribution to the optical light, and natal kick velocity}
\label{sec:distanceandveiling}
The distance to J1753 is uncertain. A most likely range of 4-8\,kpc was suggested by \citet{cadolle2007}, who compared the hydrogen column density towards the object with the total Galactic value in the line of sight. \citet{zurita2008} constrained the distance to 2.5-7.9\,kpc from the outburst X-ray light curve by calibrating its peak luminosity according to the disc instability model of \citet{king1998}. Distance estimates based on astrometric parallax and different Milky Way priors provide weak constraints of $7.15^{+3.99}_{-2.48}$\,kpc and $8.8^{+12}_{-4.0}$\,kpc \citep[][]{gandhi2019, atri2019}. Alternatively, we can quantify the distance considering the orbital period and the quiescent apparent magnitudes. For this, we employed the empirical correlation found in \citet{casares2018} between the absolute $r$-band magnitude ($M_r$) for BH XRTs during quiescence and $P_{\rm{orb}}$:
\begin{equation}
    M_{\rm{r}} = (4.64 \pm 0.10) - (3.69 \pm 0.16)\log P_{\rm{orb}} (d)~.
    \label{eq:absolutemagnitude}
\end{equation}
From $P_{\rm{orb}} = 3.26 \pm 0.02$\,h, we obtained $M_r = 7.8 \pm 0.2$. The distance was derived by comparing $M_r$ with the observed $r$-band magnitude through the modulus equation
\begin{equation}
    d (\rm{pc}) = 10^{0.2 (r - M_{\rm{r}} + 5 - A_{\rm{r}})} ~,
    \label{eq:modulusequation}
\end{equation}
where $A_r$ is the interstellar extinction in the $r$-band. We chose $A_r = 2.285 E(B-V)$ \citep{schlafly2011}, which assumes a traditional reddening law $A_V = 3.1 E(B-V)$. We used the reddening $E(B-V) = 0.45 \pm 0.09$, measured by \citet{froning2014}. This resulted in $A_r = 1.03 \pm 0.2$. On the other hand, the quiescent $r$ magnitude of J1753 can be established from the mean quiescent magnitudes $i = 21.4 \pm 0.1$ (this work) and $g = 22.62 \pm 0.1$ \citep{neustroev2017} and by applying the transformations reported by Lupton (2005).\footnote{\url{http://classic.sdss.org/dr4/algorithms/sdssUBVRITransform.html}}. Thus, we derived $r = 21.8 \pm 0.1$.\footnote{We note that while Neustroev's $g$-band magnitude was measured during early quiescence, our estimated 
$r$-band magnitude is consistent with values derived from the average of 10x30\,s $r$-band images taken between May and October 2022 by the Zwicky Transient Facility survey \citep[][$r = 21.7 \pm 0.4$]{bellm2019} as well as with a single 300 s integration on May 2021 with the 1-meter telescope at Las Cumbres Observatory ($r = 21.8 \pm 0.3$); both obtained during true quiescence. More recently, \citet{alabartarus2024} report $r = 21.7 \pm 0.2$ in full quiescence.} These numbers resulted in $d = 3.9 \pm 0.7$\,kpc. This value falls between the ranges obtained in \cite{cadolle2007} and \citet{zurita2008} and disagrees with the estimations from \citet{gandhi2019} and \cite{atri2019}.

The comparison between the absolute $r$-band magnitude obtained here ($M_r = 7.8 \pm 0.2$) with the magnitude of a M3-5 V star ($M_r = 10.9 - 13.9$)\footnote{This was calculated using \citet{covey2007} magnitudes and the transformations stated by Lupton (2005).} reflects that the optical light is dominated by the accretion flow emission. Knowing the distance, we could estimate the companion star flux contribution to the $i$-band from the photometry. The apparent $i$-band magnitude of the companion star in J1753 can be calculated by using the absolute $i$-band magnitude of the M3-5 V stars, our distance estimation, and Eq.~\ref{eq:modulusequation} with extinction $A_i = 0.76 \pm 0.2$. This led to $i$-band magnitudes in the range $23.7-26.2$, and each one has an associated uncertainty of $\pm 0.4$. The companion star fractional flux contribution to the $i$-band for J1753 in quiescence could then be quantified to be less than 15\% by obtaining the stellar to the total flux ratio. This compares well with the lower limit $8$\% obtained from the spectroscopy (Sect.~\ref{sec:averagespectralfeatures}). Using the {\sc XRBINARY} code\footnote{Software developed by E.~L. Robinson (see \url{http://www.as.utexas.edu/~elr/Robinson/XRbinary.pdf} for further details).} we computed ellipsoidal light curve models for J1753, adopting a 90\% veiling by the accretion disc, $i = 79$\,deg, $q = 0.025$, and $T_{eff} = 3210$\,K (i.e. an M4V companion). The models predict a maximum peak-to-peak amplitude in the $i$-band of $\sim 0.03$ mag, and thus we conclude that the ellipsoidal modulation is clearly masked by the strong flickering activity (0.5 mag; Sect.~\ref{sec:3.1}) and prevents its detection.

Our distance places the binary at $z = 0.8 \pm 0.2$\,kpc above the Galactic plane. The short orbital period of J1753 and its refined Galactic elevation continue to sustain the anti-correlation discovered by \citet{gandhi2020} between these two parameters. This favours the scenario of natal kicks as the mechanism responsible for positioning these BH XRTs into their current locations. The hypothesis supports the idea that these binaries originated in the Galactic disc, and a significant natal kick propelled them to higher elevations, favouring the survival of only short-period systems due to their more compact nature \citep{gandhi2020}.

In this regard, \citet{atri2019} computed the potential natal kick velocity probability distribution of J1753 based on radio astrometric proper motions and a distance of $6 \pm 2$\,kpc\, \citep{cadolle2007}. The authors ran calculations for five probable systemic radial velocity distributions with a mean from -57 to 243\,km s$^{-1}$ \citep[Table 5 in][]{atri2019} and conclude that the BH likely received a strong kick at birth. From our profiles (see Fig.\ref{fig:Tandcentroids}), we derived a mean centroid velocity of $131 \pm 55$\,km s$^{-1}$. Adopting this as the systemic velocity, we obtained a likely median kick velocity of $110^{+69}_{-37}$\,km s$^{-1}$ (90\% confidence level) when repeating the calculations of \citet{atri2019} using our distance of $3.9 \pm 0.7$\,kpc and their proper motion. However, the potential presence of a precession disc (See Sect.~\ref{sec:halfaemissionline}) can introduce long-term variability in the line centroid. This makes necessary future investigation of whether these variations are present in the H$\alpha$ profile.


\section{Conclusions}
\label{sec:5}
In this work, we have presented quiescent WHT photometry and GTC spectroscopy of J1753. The $i$-band light curve is dominated by flickering, with no signs of orbital variability. We measured a mean quiescent magnitude of $i = 21.4 \pm 0.1$ and estimated the stellar contribution to be less than 15\%. Together with the flickering, this limits the visibility of the companion star in our photometric data. However, in the average spectrum, we identified (strongly veiled) TiO bands consistent with an M4-5 V. The spectra also revealed a prominent double-peaked H$\alpha$ emission line that we used to derive constraints to the binary parameters through empirical correlations. We derived a radial velocity semi-amplitude of the companion star $K_2 = 820 \pm 36$\,km s$^{-1}$, a mass ratio $q = 0.023 \pm 0.006$, and an orbital inclination of  $i = 79 \pm 5$\,deg. Furthermore, the orbital period was determined through time analysis of the modulation detected in the depth of the trough of the H$\alpha$ line and its centroid velocities. We found an orbital period of $3.26 \pm 0.02$\,h. These numbers lead to a mass function of $f(M_1) = 7.8 \pm 1.0$\,M$_\odot$ and BH and companion star masses of $M_1 = 8.8 \pm 1.3$\,M$_\odot$ and $M_2 = 0.20 \pm 0.06$\,M$_\odot$, respectively, at the 68\% confidence level. The mass of the companion star is consistent with that of an M3-5 V, in agreement with the M4-5 V spectral classification based on TiO bands. 

Furthermore, from the quiescent magnitude we estimated a distance to J1753 of $3.9 \pm 0.7$\,kpc. This places the binary at $0.8 \pm 0.2$\,kpc above the Galactic plane. Our distance estimation supports a natal kick velocity of $110^{+69}_{-37}$\,km s$^{-1}$ for the system, although it requires confirmation.

\begin{acknowledgements}
      We thank the anonymous referee for providing useful and constructive suggestions and Pablo Rodríguez Gil for practical help during the data analysis. This article is based on observations made at the Observatorio del Roque de los Muchachos with the William Herschel Telescope (WHT) and the Gran Telescopio Canarias (GTC) operated on the island of La Palma by the IAC. IYYR, MAPT, JSS and TMD were  supported by the Agencia Estatal de Investigaci\'on del Ministerio de Ciencia e Innovaci\'on (MCIN/AEI) and the European Regional Development sFund (ERDF)  under grant PID2021-124879NB-I00, and JC via PID2022-143331NB-100. MAP acknowledges support through the Ram\'on y Cajal grant RYC2022-035388-I, funded by MCIU/AEI/10.13039/501100011033 and FSE+. PGJ is supported by the European Union (ERC, StarStruck, 101095973). Views and opinions expressed are however those of the author(s) only and do not necessarily reflect those of the European Union or the European Research Council. Neither the European Union nor the granting authority can be held responsible for them. We are thankful to the GTC staff for their excellent support during the observations.
\end{acknowledgements}

%
\bibliographystyle{aa} 
\bibliography{J1753} 

\begin{thebibliography}{65}
\expandafter\ifx\csname natexlab\endcsname\relax\def\natexlab#1{#1}\fi

\bibitem[{{Al Qasim} {et~al.}(2017){Al Qasim}, {AlMannaei}, {Russell}, {Lewis}, {Zhang}, \& {Gelfand}}]{alqasim2017}
{Al Qasim}, A., {AlMannaei}, A., {Russell}, D.~M., {et~al.} 2017, The Astronomer's Telegram, 10075, 1

\bibitem[{{Alabarta} {et~al.}(2023){Alabarta}, {Homan}, {Russell}, {Baglio}, {Saikia}, {Bramich}, {Rout}, \& {Lewis}}]{alabarta2023}
{Alabarta}, K., {Homan}, J., {Russell}, D.~M., {et~al.} 2023, The Astronomer's Telegram, 16262, 1

\bibitem[{{Alabarta} {et~al.}(2024{\natexlab{a}}){Alabarta}, {Russell}, {Bramich}, {Rout}, {Saikia}, {Baglio}, {Homan}, \& {Lewis}}]{alabartarus2024}
{Alabarta}, K., {Russell}, D.~M., {Bramich}, D.~M., {et~al.} 2024{\natexlab{a}}, The Astronomer's Telegram, 16818, 1

\bibitem[{{Alabarta} {et~al.}(2024{\natexlab{b}}){Alabarta}, {Russell}, {Bramich}, {Saikia}, {Rout}, {Baglio}, {Homan}, {Lewis}, \& {LJMU Project}}]{alabarta2024}
{Alabarta}, K., {Russell}, D.~M., {Bramich}, D.~M., {et~al.} 2024{\natexlab{b}}, The Astronomer's Telegram, 16447, 1

\bibitem[{{Atri} {et~al.}(2019){Atri}, {Miller-Jones}, {Bahramian}, {Plotkin}, {Jonker}, {Nelemans}, {Maccarone}, {Sivakoff}, {Deller}, {Chaty}, {Torres}, {Horiuchi}, {McCallum}, {Natusch}, {Phillips}, {Stevens}, \& {Weston}}]{atri2019}
{Atri}, P., {Miller-Jones}, J.~C.~A., {Bahramian}, A., {et~al.} 2019, \mnras, 489, 3116

\bibitem[{{Barthelmy} {et~al.}(2005){Barthelmy}, {Barbier}, {Cummings}, {Fenimore}, {Gehrels}, {Hullinger}, {Krimm}, {Markwardt}, {Palmer}, {Parsons}, {Sato}, {Suzuki}, {Takahashi}, {Tashiro}, \& {Tueller}}]{barthelmy2005}
{Barthelmy}, S.~D., {Barbier}, L.~M., {Cummings}, J.~R., {et~al.} 2005, \ssr, 120, 143

\bibitem[{{Bellm} {et~al.}(2019){Bellm}, {Kulkarni}, {Graham}, {Dekany}, {Smith}, {Riddle}, {Masci}, {Helou}, {Prince}, {Adams}, {Barbarino}, {Barlow}, {Bauer}, {Beck}, {Belicki}, {Biswas}, {Blagorodnova}, {Bodewits}, {Bolin}, {Brinnel}, {Brooke}, {Bue}, {Bulla}, {Burruss}, {Cenko}, {Chang}, {Connolly}, {Coughlin}, {Cromer}, {Cunningham}, {De}, {Delacroix}, {Desai}, {Duev}, {Eadie}, {Farnham}, {Feeney}, {Feindt}, {Flynn}, {Franckowiak}, {Frederick}, {Fremling}, {Gal-Yam}, {Gezari}, {Giomi}, {Goldstein}, {Golkhou}, {Goobar}, {Groom}, {Hacopians}, {Hale}, {Henning}, {Ho}, {Hover}, {Howell}, {Hung}, {Huppenkothen}, {Imel}, {Ip}, {Ivezi{\'c}}, {Jackson}, {Jones}, {Juric}, {Kasliwal}, {Kaspi}, {Kaye}, {Kelley}, {Kowalski}, {Kramer}, {Kupfer}, {Landry}, {Laher}, {Lee}, {Lin}, {Lin}, {Lunnan}, {Giomi}, {Mahabal}, {Mao}, {Miller}, {Monkewitz}, {Murphy}, {Ngeow}, {Nordin}, {Nugent}, {Ofek}, {Patterson}, {Penprase}, {Porter}, {Rauch}, {Rebbapragada}, {Reiley}, {Rigault}, {Rodriguez}, {van Roestel}, {Rusholme}, {van
  Santen}, {Schulze}, {Shupe}, {Singer}, {Soumagnac}, {Stein}, {Surace}, {Sollerman}, {Szkody}, {Taddia}, {Terek}, {Van Sistine}, {van Velzen}, {Vestrand}, {Walters}, {Ward}, {Ye}, {Yu}, {Yan}, \& {Zolkower}}]{bellm2019}
{Bellm}, E.~C., {Kulkarni}, S.~R., {Graham}, M.~J., {et~al.} 2019, \pasp, 131, 018002

\bibitem[{{Benn} {et~al.}(2008){Benn}, {Dee}, \& {Ag{\'o}cs}}]{benn2008}
{Benn}, C., {Dee}, K., \& {Ag{\'o}cs}, T. 2008, in Society of Photo-Optical Instrumentation Engineers (SPIE) Conference Series, Vol. 7014, Ground-based and Airborne Instrumentation for Astronomy II, ed. I.~S. {McLean} \& M.~M. {Casali}, 70146X

\bibitem[{{Bernardini} {et~al.}(2017){Bernardini}, {Zhang}, {Russell}, {Gelfand}, {Qasim}, {AlMannaei}, {Lewis}, {Shaw}, {Tomsick}, \& {Plotkin}}]{bernardini2017}
{Bernardini}, F., {Zhang}, G., {Russell}, D.~M., {et~al.} 2017, The Astronomer's Telegram, 10325, 1

\bibitem[{{Cadolle Bel} {et~al.}(2007){Cadolle Bel}, {Rib{\'o}}, {Rodriguez}, {Chaty}, {Corbel}, {Goldwurm}, {Frontera}, {Farinelli}, {D'Avanzo}, {Tarana}, {Ubertini}, {Laurent}, {Goldoni}, \& {Mirabel}}]{cadolle2007}
{Cadolle Bel}, M., {Rib{\'o}}, M., {Rodriguez}, J., {et~al.} 2007, \apj, 659, 549

\bibitem[{{Casares}(2015)}]{casares2015}
{Casares}, J. 2015, \apj, 808, 80

\bibitem[{{Casares}(2016)}]{Casares2016}
{Casares}, J. 2016, \apj, 822, 99

\bibitem[{{Casares}(2018)}]{casares2018}
{Casares}, J. 2018, \mnras, 473, 5195

\bibitem[{{Casares} \& {Jonker}(2014)}]{casares2014}
{Casares}, J. \& {Jonker}, P.~G. 2014, \ssr, 183, 223

\bibitem[{{Casares} {et~al.}(2022){Casares}, {Mu{\~n}oz-Darias}, {Torres}, {Mata S{\'a}nchez}, {Britt}, {Armas Padilla}, {{\'A}lvarez-Hern{\'a}ndez}, {C{\'u}neo}, {Gonz{\'a}lez Hern{\'a}ndez}, {Jim{\'e}nez-Ibarra}, {Jonker}, {Panizo-Espinar}, {S{\'a}nchez-Sierras}, \& {Yanes-Rizo}}]{casares2022}
{Casares}, J., {Mu{\~n}oz-Darias}, T., {Torres}, M.~A.~P., {et~al.} 2022, \mnras, 516, 2023

\bibitem[{{Cepa} {et~al.}(2000){Cepa}, {Aguiar}, {Escalera}, {Gonzalez-Serrano}, {Joven-Alvarez}, {Peraza}, {Rasilla}, {Rodriguez-Ramos}, {Gonzalez}, {Cobos Duenas}, {Sanchez}, {Tejada}, {Bland-Hawthorn}, {Militello}, \& {Rosa}}]{cepa2000}
{Cepa}, J., {Aguiar}, M., {Escalera}, V.~G., {et~al.} 2000, in Society of Photo-Optical Instrumentation Engineers (SPIE) Conference Series, Vol. 4008, Optical and IR Telescope Instrumentation and Detectors, ed. M.~{Iye} \& A.~F. {Moorwood}, 623--631

\bibitem[{{Chambers} {et~al.}(2016){Chambers}, {Magnier}, {Metcalfe}, {Flewelling}, {Huber}, {Waters}, {Denneau}, {Draper}, {Farrow}, {Finkbeiner}, {Holmberg}, {Koppenhoefer}, {Price}, {Rest}, {Saglia}, {Schlafly}, {Smartt}, {Sweeney}, {Wainscoat}, {Burgett}, {Chastel}, {Grav}, {Heasley}, {Hodapp}, {Jedicke}, {Kaiser}, {Kudritzki}, {Luppino}, {Lupton}, {Monet}, {Morgan}, {Onaka}, {Shiao}, {Stubbs}, {Tonry}, {White}, {Ba{\~n}ados}, {Bell}, {Bender}, {Bernard}, {Boegner}, {Boffi}, {Botticella}, {Calamida}, {Casertano}, {Chen}, {Chen}, {Cole}, {Deacon}, {Frenk}, {Fitzsimmons}, {Gezari}, {Gibbs}, {Goessl}, {Goggia}, {Gourgue}, {Goldman}, {Grant}, {Grebel}, {Hambly}, {Hasinger}, {Heavens}, {Heckman}, {Henderson}, {Henning}, {Holman}, {Hopp}, {Ip}, {Isani}, {Jackson}, {Keyes}, {Koekemoer}, {Kotak}, {Le}, {Liska}, {Long}, {Lucey}, {Liu}, {Martin}, {Masci}, {McLean}, {Mindel}, {Misra}, {Morganson}, {Murphy}, {Obaika}, {Narayan}, {Nieto-Santisteban}, {Norberg}, {Peacock}, {Pier}, {Postman}, {Primak}, {Rae}, {Rai},
  {Riess}, {Riffeser}, {Rix}, {R{\"o}ser}, {Russel}, {Rutz}, {Schilbach}, {Schultz}, {Scolnic}, {Strolger}, {Szalay}, {Seitz}, {Small}, {Smith}, {Soderblom}, {Taylor}, {Thomson}, {Taylor}, {Thakar}, {Thiel}, {Thilker}, {Unger}, {Urata}, {Valenti}, {Wagner}, {Walder}, {Walter}, {Watters}, {Werner}, {Wood-Vasey}, \& {Wyse}}]{chambers2016}
{Chambers}, K.~C., {Magnier}, E.~A., {Metcalfe}, N., {et~al.} 2016, arXiv e-prints, arXiv:1612.05560

\bibitem[{{Corral-Santana} {et~al.}(2016){Corral-Santana}, {Casares}, {Mu{\~n}oz-Darias}, {Bauer}, {Mart{\'\i}nez-Pais}, \& {Russell}}]{corral2016}
{Corral-Santana}, J.~M., {Casares}, J., {Mu{\~n}oz-Darias}, T., {et~al.} 2016, \aap, 587, A61

\bibitem[{{Corral-Santana} {et~al.}(2018){Corral-Santana}, {Torres}, {Shahbaz}, {Bartlett}, {Russell}, {Kong}, {Casares}, {Mu{\~n}oz-Darias}, {Bauer}, {Homan}, {Jonker}, {Mata S{\'a}nchez}, {Wevers}, {Rodr{\'\i}guez-Gil}, {Lewis}, \& {Schreuder}}]{corral2018}
{Corral-Santana}, J.~M., {Torres}, M. A.~P., {Shahbaz}, T., {et~al.} 2018, \mnras, 475, 1036

\bibitem[{{Covey} {et~al.}(2007){Covey}, {Ivezi{\'c}}, {Schlegel}, {Finkbeiner}, {Padmanabhan}, {Lupton}, {Ag{\"u}eros}, {Bochanski}, {Hawley}, {West}, {Seth}, {Kimball}, {Gogarten}, {Claire}, {Haggard}, {Kaib}, {Schneider}, \& {Sesar}}]{covey2007}
{Covey}, K.~R., {Ivezi{\'c}}, {\v{Z}}., {Schlegel}, D., {et~al.} 2007, \aj, 134, 2398

\bibitem[{{C{\'u}neo} {et~al.}(2023){C{\'u}neo}, {Casares}, {Armas Padilla}, {S{\'a}nchez-Sierras}, {Corral-Santana}, {Maccarone}, {Mata S{\'a}nchez}, {Mu{\~n}oz-Darias}, {Torres}, \& {Vincentelli}}]{cuneo2023}
{C{\'u}neo}, V.~A., {Casares}, J., {Armas Padilla}, M., {et~al.} 2023, \aap, 679, L11

\bibitem[{{Durant} {et~al.}(2009){Durant}, {Gandhi}, {Shahbaz}, {Peralta}, \& {Dhillon}}]{durant2009}
{Durant}, M., {Gandhi}, P., {Shahbaz}, T., {Peralta}, H.~H., \& {Dhillon}, V.~S. 2009, \mnras, 392, 309

\bibitem[{{Fender} {et~al.}(2005){Fender}, {Garrington}, \& {Muxlow}}]{fender2005}
{Fender}, R., {Garrington}, S., \& {Muxlow}, T. 2005, The Astronomer's Telegram, 558, 1

\bibitem[{{Frank} {et~al.}(2002){Frank}, {King}, \& {Raine}}]{frank2002}
{Frank}, J., {King}, A., \& {Raine}, D.~J. 2002, {Accretion Power in Astrophysics: Third Edition}

\bibitem[{{Froning} {et~al.}(2014){Froning}, {Maccarone}, {France}, {Winter}, {Robinson}, {Hynes}, \& {Lewis}}]{froning2014}
{Froning}, C.~S., {Maccarone}, T.~J., {France}, K., {et~al.} 2014, \apj, 780, 48

\bibitem[{{Gandhi} {et~al.}(2020){Gandhi}, {Rao}, {Charles}, {Belczynski}, {Maccarone}, {Arur}, \& {Corral-Santana}}]{gandhi2020}
{Gandhi}, P., {Rao}, A., {Charles}, P.~A., {et~al.} 2020, \mnras, 496, L22

\bibitem[{{Gandhi} {et~al.}(2019){Gandhi}, {Rao}, {Johnson}, {Paice}, \& {Maccarone}}]{gandhi2019}
{Gandhi}, P., {Rao}, A., {Johnson}, M. A.~C., {Paice}, J.~A., \& {Maccarone}, T.~J. 2019, \mnras, 485, 2642

\bibitem[{{Halpern}(2005)}]{halpern2005}
{Halpern}, J.~P. 2005, The Astronomer's Telegram, 549, 1

\bibitem[{{Jonker} {et~al.}(2008){Jonker}, {Torres}, \& {Steeghs}}]{jonker2008}
{Jonker}, P.~G., {Torres}, M.~A.~P., \& {Steeghs}, D. 2008, in American Institute of Physics Conference Series, Vol. 1010, A Population Explosion: The Nature \& Evolution of X-ray Binaries in Diverse Environments, ed. R.~M. {Bandyopadhyay}, S.~{Wachter}, D.~{Gelino}, \& C.~R. {Gelino} (AIP), 109--116

\bibitem[{{King} \& {Ritter}(1998)}]{king1998}
{King}, A.~R. \& {Ritter}, H. 1998, \mnras, 293, L42

\bibitem[{{Lomb}(1976)}]{lomb1976}
{Lomb}, N.~R. 1976, \apss, 39, 447

\bibitem[{{Marsh}(1990)}]{marsh1999}
{Marsh}, T.~R. 1990, \apj, 357, 621

\bibitem[{{Mata S{\'a}nchez} {et~al.}(2015){Mata S{\'a}nchez}, {Mu{\~n}oz-Darias}, {Casares}, {Corral-Santana}, \& {Shahbaz}}]{mata2015}
{Mata S{\'a}nchez}, D., {Mu{\~n}oz-Darias}, T., {Casares}, J., {Corral-Santana}, J.~M., \& {Shahbaz}, T. 2015, \mnras, 454, 2199

\bibitem[{{McClintock} \& {Remillard}(2006)}]{mcclintock2006}
{McClintock}, J.~E. \& {Remillard}, R.~A. 2006, in Compact stellar X-ray sources, Vol.~39, 157--213

\bibitem[{{Naylor}(1998)}]{Naylor1998}
{Naylor}, T. 1998, \mnras, 296, 339

\bibitem[{{Neustroev} {et~al.}(2017){Neustroev}, {Zharikov}, \& {Cabrera-Lavers}}]{neustroev2017}
{Neustroev}, V., {Zharikov}, S., \& {Cabrera-Lavers}, A. 2017, The Astronomer's Telegram, 10664, 1

\bibitem[{{Neustroev} {et~al.}(2014){Neustroev}, {Veledina}, {Poutanen}, {Zharikov}, {Tsygankov}, {Sjoberg}, \& {Kajava}}]{neustroev2014}
{Neustroev}, V.~V., {Veledina}, A., {Poutanen}, J., {et~al.} 2014, \mnras, 445, 2424

\bibitem[{{Orosz} {et~al.}(1994){Orosz}, {Bailyn}, {Remillard}, {McClintock}, \& {Foltz}}]{orosz1994}
{Orosz}, J.~A., {Bailyn}, C.~D., {Remillard}, R.~A., {McClintock}, J.~E., \& {Foltz}, C.~B. 1994, \apj, 436, 848

\bibitem[{{Paez} {et~al.}(2012){Paez}, {Mason}, {Robinson}, \& {Bayless}}]{paez2012}
{Paez}, A., {Mason}, P.~A., {Robinson}, E.~L., \& {Bayless}, A.~J. 2012, in American Astronomical Society Meeting Abstracts, Vol. 219, American Astronomical Society Meeting Abstracts \#219, 153.07

\bibitem[{{Palmer} {et~al.}(2005){Palmer}, {Barthelmey}, {Cummings}, {Gehrels}, {Krimm}, {Markwardt}, {Sakamoto}, \& {Tueller}}]{palmer2005}
{Palmer}, D.~M., {Barthelmey}, S.~D., {Cummings}, J.~R., {et~al.} 2005, The Astronomer's Telegram, 546, 1

\bibitem[{{Pecaut} \& {Mamajek}(2013)}]{pecaut2013}
{Pecaut}, M.~J. \& {Mamajek}, E.~E. 2013, \apjs, 208, 9

\bibitem[{{Rahoui} {et~al.}(2015){Rahoui}, {Tomsick}, {Coriat}, {Corbel}, {F{\"u}rst}, {Gandhi}, {Kalemci}, {Migliari}, {Stern}, \& {Tzioumis}}]{rahoui2015}
{Rahoui}, F., {Tomsick}, J.~A., {Coriat}, M., {et~al.} 2015, \apj, 810, 161

\bibitem[{{Russell} {et~al.}(2016){Russell}, {AlMannaei}, {Qasim}, {Shaw}, {Charles}, \& {Lewis}}]{russel2016}
{Russell}, D.~M., {AlMannaei}, A., {Qasim}, A.~A., {et~al.} 2016, The Astronomer's Telegram, 9708, 1

\bibitem[{{S{\'a}nchez-Sierras} {et~al.}(2023){S{\'a}nchez-Sierras}, {Mu{\~n}oz-Darias}, {Casares}, {Panizo-Espinar}, {Armas Padilla}, {Corral-Santana}, {C{\'u}neo}, {Mata S{\'a}nchez}, {Motta}, {Ponti}, {Steeghs}, {Torres}, \& {Vincentelli}}]{sanchez2023}
{S{\'a}nchez-Sierras}, J., {Mu{\~n}oz-Darias}, T., {Casares}, J., {et~al.} 2023, \aap, 673, A104

\bibitem[{{Scargle}(1982)}]{scargle1982}
{Scargle}, J.~D. 1982, \apj, 263, 835

\bibitem[{{Schlafly} \& {Finkbeiner}(2011)}]{schlafly2011}
{Schlafly}, E.~F. \& {Finkbeiner}, D.~P. 2011, \apj, 737, 103

\bibitem[{{Shahbaz} {et~al.}(2013){Shahbaz}, {Russell}, {Zurita}, {Casares}, {Corral-Santana}, {Dhillon}, \& {Marsh}}]{shahbaz2013}
{Shahbaz}, T., {Russell}, D.~M., {Zurita}, C., {et~al.} 2013, \mnras, 434, 2696

\bibitem[{{Shaw} {et~al.}(2016){Shaw}, {Charles}, {Casares}, \& {Hern{\'a}ndez Santisteban}}]{shaw2016}
{Shaw}, A.~W., {Charles}, P.~A., {Casares}, J., \& {Hern{\'a}ndez Santisteban}, J.~V. 2016, \mnras, 463, 1314

\bibitem[{{Smette} {et~al.}(2015){Smette}, {Sana}, {Noll}, {Horst}, {Kausch}, {Kimeswenger}, {Barden}, {Szyszka}, {Jones}, {Gallenne}, {Vinther}, {Ballester}, \& {Taylor}}]{smette2015}
{Smette}, A., {Sana}, H., {Noll}, S., {et~al.} 2015, \aap, 576, A77

\bibitem[{{Soleri} {et~al.}(2013){Soleri}, {Mu{\~n}oz-Darias}, {Motta}, {Belloni}, {Casella}, {M{\'e}ndez}, {Altamirano}, {Linares}, {Wijnands}, {Fender}, \& {van der Klis}}]{soleri2013}
{Soleri}, P., {Mu{\~n}oz-Darias}, T., {Motta}, S., {et~al.} 2013, \mnras, 429, 1244

\bibitem[{{Still} {et~al.}(2005){Still}, {Roming}, {Brocksopp}, \& {Markwardt}}]{still2005}
{Still}, M., {Roming}, P., {Brocksopp}, C., \& {Markwardt}, C.~B. 2005, The Astronomer's Telegram, 553, 1

\bibitem[{{Torres} {et~al.}(2004){Torres}, {Callanan}, {Garcia}, {Zhao}, {Laycock}, \& {Kong}}]{torres2004}
{Torres}, M.~A.~P., {Callanan}, P.~J., {Garcia}, M.~R., {et~al.} 2004, \apj, 612, 1026

\bibitem[{{Torres} {et~al.}(2015){Torres}, {Jonker}, {Miller-Jones}, {Steeghs}, {Repetto}, \& {Wu}}]{torres2015}
{Torres}, M.~A.~P., {Jonker}, P.~G., {Miller-Jones}, J.~C.~A., {et~al.} 2015, \mnras, 450, 4292

\bibitem[{{Torres} {et~al.}(2005{\natexlab{a}}){Torres}, {Steeghs}, {Blake}, {Jonker}, {Garcia}, {McClintock}, {Miller}, {Zhao}, {Calkins}, {Berlind}, {Falco}, {Bloom}, {Callanan}, \& {Rodriguez-Gil}}]{torres2005b}
{Torres}, M.~A.~P., {Steeghs}, D., {Blake}, C., {et~al.} 2005{\natexlab{a}}, The Astronomer's Telegram, 566, 1

\bibitem[{{Torres} {et~al.}(2005{\natexlab{b}}){Torres}, {Steeghs}, {Garcia}, {McClintock}, {Miller}, {Jonker}, {Callanan}, {Zhao}, {Huchra}, {U}, \& {Hutcheson}}]{torres2005}
{Torres}, M.~A.~P., {Steeghs}, D., {Garcia}, M.~R., {et~al.} 2005{\natexlab{b}}, The Astronomer's Telegram, 551, 1

\bibitem[{{Vande Putte} {et~al.}(2003){Vande Putte}, {Smith}, {Hawkins}, \& {Martin}}]{vandeputte2003}
{Vande Putte}, D., {Smith}, R.~C., {Hawkins}, N.~A., \& {Martin}, J.~S. 2003, \mnras, 342, 151

\bibitem[{{Veledina} {et~al.}(2015){Veledina}, {Revnivtsev}, {Durant}, {Gandhi}, \& {Poutanen}}]{veledina2015}
{Veledina}, A., {Revnivtsev}, M.~G., {Durant}, M., {Gandhi}, P., \& {Poutanen}, J. 2015, \mnras, 454, 2855

\bibitem[{{Verro} {et~al.}(2022){Verro}, {Trager}, {Peletier}, {Lan{\c{c}}on}, {Gonneau}, {Vazdekis}, {Prugniel}, {Chen}, {Coelho}, {S{\'a}nchez-Bl{\'a}zquez}, {Martins}, {Arentsen}, {Lyubenova}, {Falc{\'o}n-Barroso}, \& {Dries}}]{verro2022}
{Verro}, K., {Trager}, S.~C., {Peletier}, R.~F., {et~al.} 2022, \aap, 660, A34

\bibitem[{{Wade} \& {Horne}(1988)}]{wade1988}
{Wade}, R.~A. \& {Horne}, K. 1988, \apj, 324, 411

\bibitem[{{Yanes-Rizo} {et~al.}(2024){Yanes-Rizo}, {Torres}, {Casares}, {Monelli}, {Jonker}, {Abbot}, {Armas Padilla}, \& {Mu{\~n}oz-Darias}}]{yanes2024}
{Yanes-Rizo}, I.~V., {Torres}, M.~A.~P., {Casares}, J., {et~al.} 2024, \mnras, 527, 5949

\bibitem[{{Yang} {et~al.}(2022){Yang}, {Zhang}, {Russell}, {Gelfand}, {M{\'e}ndez}, {Wang}, \& {Lyu}}]{yang2022}
{Yang}, P., {Zhang}, G., {Russell}, D.~M., {et~al.} 2022, \mnras, 514, 234

\bibitem[{{Yoshikawa} {et~al.}(2015){Yoshikawa}, {Yamada}, {Nakahira}, {Matsuoka}, {Negoro}, {Mihara}, \& {Tamagawa}}]{yoshikawa2015}
{Yoshikawa}, A., {Yamada}, S., {Nakahira}, S., {et~al.} 2015, \pasj, 67, 11

\bibitem[{{Zhang} {et~al.}(2017){Zhang}, {Russell}, {Bernardini}, {Gelfand}, \& {Lewis}}]{zhang2017a}
{Zhang}, G., {Russell}, D.~M., {Bernardini}, F., {Gelfand}, J.~D., \& {Lewis}, F. 2017, The Astronomer's Telegram, 10562, 1

\bibitem[{{Zurita} {et~al.}(2002){Zurita}, {Casares}, {Shahbaz}, {Wagner}, {Foltz}, {Rodr{\'\i}guez-Gil}, {Hynes}, {Charles}, {Ryan}, {Schwarz}, \& {Starrfield}}]{zurita2002}
{Zurita}, C., {Casares}, J., {Shahbaz}, T., {et~al.} 2002, \mnras, 333, 791

\bibitem[{{Zurita} {et~al.}(2008){Zurita}, {Durant}, {Torres}, {Shahbaz}, {Casares}, \& {Steeghs}}]{zurita2008}
{Zurita}, C., {Durant}, M., {Torres}, M.~A.~P., {et~al.} 2008, \apj, 681, 1458

\end{thebibliography}
%

\end{document}